\documentclass[useAMS,usenatbib,usegraphicx,iop,appendixfloats,twocolumn]{aastex62}

\usepackage{verbatim}
\usepackage{xspace}
\usepackage{rotating}
\usepackage{array}
\usepackage{url}

\def\micron{$\mu$m\xspace}
\def\htwo{H$_{2}$\xspace}

\def\hi{H$\,$\textsc{i}\xspace}
\def\ci{[C$\,$\textsc{i}]\xspace}

\def\cii{[C$\,$\textsc{ii}]\xspace}
\def\nii{[N$\,$\textsc{ii}]\xspace}

\def\cione{[C$\,$\textsc{i}](1-0)\xspace}
\def\citwo{[C$\,$\textsc{i}](2-1)\xspace}

\def\twco{$^{12}$CO\xspace}

\def\alphaCO{$\alpha_{\mathrm{CO}}$\xspace}
\def\alphaCOone{$\alpha_{\mathrm{CO(1-0)}}$\xspace}
\def\alphaCOtwo{$\alpha_{\mathrm{CO(2-1)}}$\xspace}
\def\alphaCI{$\alpha_{\mathrm{[CI]}}$\xspace}
\def\alphaCIone{$\alpha_{\mathrm{[CI](1-0)}}$\xspace}
\def\alphaCItwo{$\alpha_{\mathrm{[CI](2-1)}}$\xspace}

\shorttitle{\cione and \citwo in resolved local galaxies}

\begin{document}

\title{\cione and \citwo in resolved local galaxies\footnote{{\it Herschel} is an ESA space observatory with science instruments provided by European-led Principal Investigator consortia and with important participation from NASA.}}

\author{Alison F. Crocker}
\affiliation{Department of Physics, Reed College, Portland, OR, 97202}

\author{Eric Pellegrini}
\affiliation{University of Heidelberg Institute for Theoretical Astrophysics, 69120 Heidelberg,
Germany}

\author{J.-D. T. Smith}
\affiliation{Department of Physics and Astronomy, University of Toledo, Toledo, OH, 43606}

\author{Bruce T. Draine}
\affiliation{Princeton University Observatory, Peyton Hall, Princeton, NJ 08544-1001, USA}

\author{Christine D. Wilson}
\affiliation{Department of Physics \& Astronomy, McMaster University, Hamilton, Ontario, Canada}

\author{Mark Wolfire}
\affiliation{Department of Astronomy, University of Maryland, College Park, MD 20742-2421, USA}

\author{Lee Armus}
\affiliation{IPAC, California Institute of Technology, Pasadena, CA 91125, USA}

\author{ Elias Brinks}
\affiliation{Centre for Astrophysics Research, University of Hertfordshire, College Lane, Hatfield, AL10 9AB, UK}

\author{Daniel A. Dale}
\affiliation{Department of Physics \& Astronomy, University of Wyoming, Laramie WY, USA}

\author{ Brent Groves}
\affiliation{Research School of Astronomy \& Astrophysics, Australian National University, Canberra, Australia}

\author{Rodrigo Herrera-Camus}
\affiliation{Departamento de Astronom\'{i}a, Facultad de Ciencias F\'{i}sicas y Matematicas, Universidad de Concepci\'{o}n, Avenida Esteban Iturra s/n, Casilla 160-C, Concepci\'{o}n, Chile}

\author{ Leslie K. Hunt}
\affiliation{INAF-Osservatorio Astrofisico di Arcetri, Largo E. Fermi, 5, 50125, Firenze, Italy}

\author{Robert C. Kennicutt}
\affiliation{Institute of Astronomy, University of Cambridge, Cambridge, UK}

\author{Eric J. Murphy}
\affiliation{National Radio Astronomy Observatory, Charlottesville, VA, USA}

\author{Karin Sandstrom}
\affiliation{Center for Astrophysics and Space Science, University of California, San Diego CA, USA}

\author{ Eva Schinnerer}
\affiliation{MPI for Astronomy, K\"{o}nigstuhl 17, D-69117, Heidelberg, Germany}

\author{Dimitra Rigopoulou}
\affiliation{Department of Physics, University of Oxford, Keble Road, Oxford OX1 3RH, UK}

\author{Erik Rosolowsky}
\affiliation{University of British Columbia Okanagan, 3333 University Way, Kelowna, BC V1V 1V7}

\author{ Paul van der Werf}
\affiliation{Leiden Observatory, Leiden University, P.O. Box 9513, NL-2300 RA Leiden, The Netherlands}

\begin{abstract}   
We present resolved \ci line intensities of 18 nearby galaxies observed with the SPIRE FTS spectrometer on the Herschel Space Observatory. We use these data along with resolved CO line intensities from $J_\mathrm{up} = 1$ to $7$ to interpret what phase of the interstellar medium the \ci lines trace within typical local galaxies. A tight, linear relation is found between the intensities of the CO(4-3) and \citwo lines; we hypothesize this is due to the similar upper level temperature of these two lines. 
We modeled the \ci and CO line emission using large velocity gradient models
combined with an empirical template. According to this modeling, the \cione line is clearly dominated by the low-excitation component.
We determine \ci to molecular mass conversion factors for both the \cione and \citwo lines, with mean values of $\alpha_{\mathrm{[CI](1-0)}} = 7.3$ M$_{\sun}$ K$^{-1}$ km$^{-1}$ s pc$^{-2}$ and $\alpha_{\mathrm{[CI](2-1)}} = 34 $ M$_{\sun}$ K$^{-1}$ km$^{-1}$ s pc$^{-2}$ with logarithmic root-mean-square spreads of 0.20 and 0.32 dex, respectively.  The similar spread of \alphaCIone to \alphaCO (derived using the CO(2-1) line) suggests that \cione may be just as good a tracer of cold molecular gas as CO(2-1) in galaxies of this type. On the other hand, the wider spread of $\alpha_{\mathrm{[CI](2-1)}}$ and the tight relation found between \citwo and CO(4-3) suggest that much of the \citwo emission may originate in warmer molecular gas. \end{abstract}

\section{Introduction}

In this paper, we study the resolved \ci line emission obtained by the Herschel Space Observatory for a sample of nearby (mostly disk) galaxies. We focus on the potential diagnostic power of the \ci lines and, in particular, whether they may be used as tracers of the total molecular gas content.

Ground state neutral carbon (C$^{0}$) emits two fine-structure emission lines in the far-infrared (FIR). The \cione and \citwo lines have wavelengths of 609 and 370 \micron, frequencies of 492 and 809 GHz, and upper level temperatures of 23.6 and 62.4 K, respectively. Given its low ionization potential of 11.26~eV, C$^{0}$ is found mostly in the cold neutral and molecular ISM and within photodissociation regions (PDRs, e.g., \citet{tielens85}). For static and homogenous PDRs, the neutral carbon is expected to reside in a layer between C$^{+}$ and CO at a range of extinctions into the cloud from the dissociating source of $A_{\mathrm V} \approx 2-4$ \citep{tielens85}. In more realistic molecular clouds with clumps, turbulence and possibly embedded stellar sources, the neutral carbon `layer' is a complicated surface that may effectively exist throughout the cloud \citep{offner14, glover15}.  This distribution is directly seen in observations of Milky Way molecular clouds \citep{shimajiri13}. 

The \ci \xspace lines are more easily observed at high redshift and thus have already been used as ISM diagnostics \citep[e.g.][]{walter11, carilli13}. However, much of the low redshift-observational work that grounds these interpretations is limited to the Milky Way or has thus far focused on high-excitation galaxies, such as LIRGs. Using Herschel Space Observatory data, \citet{israel15} (hereafter Is15) studied a group of 76 LIRGs and starburst galaxy centers in \ci  and CO lines. They establish many empirical correlations between \ci and CO line intensities and between ratios of these lines.  Then they perform large velocity gradient (LVG) modeling of three fiducial sets of \ci  and CO ratios in order to describe different families of galaxies within their sample. 

\citet{kamen16} published a master catalog of central CO, \ci and \nii\xspace line fluxes for 301 extragalactic sources observed with Herschel. However, these central-only measurements do not indicate the typical ISM properties within galaxy disks, which require resolved observations. Two papers from the Very Nearby Galaxy Survey present and analyze CO and \ci \xspace maps from Herschel SPIRE spectroscopy of nearby spirals M83 \citep{wu15} and M51 \citep{schirm17}. \citet{wu15} reports an interesting linear relationship between \citwo \xspace intensity and the surface density of star formation. \citet{schirm17} are able to include the \ci \xspace lines along with the CO lines to fit a two-component non-LTE model describing the state of the molecular gas in M51. They find the cold component is similar to that found in more excited galaxies and the warm component has a similar temperature, but is less dense.  Very recently, a paper with a similar sample to ours (nine galaxies overlap) published maps of the \ci \xspace lines for 15 local spiral galaxies \citep{jiao19} and studies the correlations of \ci \xspace  lines with CO(1-0) and dust emission.

Extragalactic molecular gas measurements typically depend on \twco line emission to obtain molecular masses through a conversion factor $\alpha_{\mathrm{CO}}$: $M_{\mathrm{mol}} = \alpha_{\mathrm{CO}} L_{\mathrm{CO}}$, which is known to vary with certain galaxy properties (metallicity, IR luminosity; see \citet{bolatto13} for a recent review). 
The \ci emission has also received attention as a potential molecular gas tracer, particularly as one that might excel at higher redshift. 
Comparison of the \cione line to various CO lines and the dust continuum within 14 nearby galaxies led \citet{gerin00} to conclude that \cione was a preferred tracer of the H$_{2}$ mass compared to the CO lines. \citet{papadopoulos04} present non-equilibrium chemical models and suggest that this non-equilibrium chemistry and the known turbulence of molecular clouds keeps the abundance of neutral carbon high throughout molecular clouds. They thus argue that \ci should be a good tracer of the total molecular gas, especially highlighting how the \ci lines may be helpful at high redshift where they are easier to observe than low-$J$ CO lines.
 
 \citet{offner14} and \citet{glover15} simulated individual, turbulent molecular clouds and focused on their \ci content and emission. They both find that \ci traces the column density of H$_{2}$ in their simulated clouds, with a conversion factor of $X_{\mathrm{[CI](1-0)}} = N_{\mathrm H_{2}}/I_{\mathrm{[CI](1-0)}} = 1.0-1.1 \times 10^{21}$ cm$^{-2}$ K$^{-1}$ km$^{-1}$ s (corresponding to a molecular mass conversion factor of $\alpha_{\mathrm{[CI](1-0)}} \approx 20$ M$_{\sun}$ K$^{-1}$ km$^{-1}$ s pc$^{-2}$, including helium). In both cases, the simulations did not include radiation from newly-formed internal stellar sources. So, in particular, their simulations miss any contribution from PDRs irradiated by high-intensity radiation fields. A more recent simulation by \citet{clark19} shows that \ci (particularly \cione) traces molecular gas with similar characteristics as CO(1-0): gas with number densities 500-1000 cm$^{-3}$ and kinetic temperatures under 30K.

Recent work on local galaxies has analyzed the utility of \cione and \citwo \xspace as molecular gas tracers, with conflicting results. \citet{israel15} recommend against using either \ci line as a molecular gas mass tracer having found a poor correlation between the \ci lines and their beam-averaged $H_2$ column density. In contrast, \citet{jiao17} also use Herschel data on LIRGs along with CO(1-0) data to argue that, due to a strong (linear) relation between the \ci and CO(1-0) line luminosities, the \ci lines can be used to measure the total H$_{2}$ mass of these galaxies. Most recently, \citet{jiao19} extend this to local spiral galaxies and again conclude the \cione \xspace is likely a good molecular mass tracer based upon its good correlation with CO(1-0).

Our paper focuses on the resolved \ci emission in a sample of local galaxies. After presenting the sample and data in Section 2, we first report an empirical analysis of the \ci lines (Section 3). In order to understand what physical properties are driving the relationships we see, we use models applied to the \ci and full suite of CO lines from $J_\mathrm{up} = 1$ to $7$ in Section 4. Finally, we conclude with an investigation of the H$_{2}$ conversion factor for the \ci lines, $\alpha_{\mathrm{[CI]}}$, in Section 5.

\section{Data}

\subsection{Beyond the Peak data}

The Beyond the Peak (BtP) project used the SPIRE spectrometer \citep{griffin10} on the Herschel Space Observatory \citep{pilbratt10} to target 22 local star-forming galaxies taken from the KINGFISH project sample (see Table~1). Galaxies were chosen to represent a range of gas conditions, star formation rates and galaxy masses while maintaining a central infrared surface brightness of $9 \times 10^{-6}$ W m$^{-2}$ sr$^{-1}$ or brighter in a 40\arcsec \xspace aperture. Table~1 documents the physical resolution achieved for each galaxy and whether each galaxy's \ci emission is not detected (below a S/N of 3), detected only in the center or detected in multiple resolution elements (resolved). Note that the physical resolution is computed based upon 40\arcsec \xspace extraction apertures; the actual spatial resolution is either 37\arcsec~for the longer-wavelength SLW bolometer data and 19\arcsec~for the shorter-wavelength SSW bolometer data. Further properties of these galaxies such as distance, metallicity, stellar mass and star formation rate are documented in Table~1 of \citet{kennicutt11}.

The Herschel SPIRE FTS data were obtained and reduced as explained in detail in \citet{pellegrini13}. To more reliably extract the faint, extended emission, many custom modifications were made to the Herschel pipeline. A semi-extended source correction was applied to Level 1 data, after which spectra were extracted from 40\arcsec\xspace diameter regions. We choose to extract on a rectilinear grid with beam centers separated by 40\arcsec\xspace  so that the regions are statistically independent. We then fitted a modified black-body with a scalar offset to each SPIRE/FTS spectrum (omitting ranges with line emission) and removed this offset. Synthetic photometry in the three SPIRE photometric bands was then computed from these offset-corrected spectra. We flux calibrated the spectra by using a scale factor to match this synthetic photometry to that measured by the SPIRE photometer in identical apertures. 
After this flux calibration was applied, line fluxes were extracted by fitting gauss-sinc profiles to the FTS spectra. This profile is appropriate because even at the highest spectral resolution at short wavelengths (300 km s$^{-1}$) nearly all the galaxies had unresolved spectral lines. The one exception was NGC~7331, which required convolution with a gaussian profile due to physical spectral broadening from an inclined star forming ring in the beam. Spectral lines were considered detected if they were above 3$\sigma$. The line intensities for \cione and \citwo \xspace  are tabulated in Table~\ref{tab:cidata} for all 40\arcsec\xspace diameter regions. The CO line intensities will be presented in a subsequent paper.

\begin{table}[htp]
\begin{center}
\caption{BtP sample galaxies: basic parameters and coverage.}
\label{tab:sample}
\begin{tabular}{lccccc}
\tableline
Galaxy & 40'' Res.& $T_{\mathrm{d}}$ (K) &  $T_{\mathrm{d}}$ (K) & Enc.  & Frac. \\
 & (kpc) & central & disk & 70\micron  & $R_{25}$\\
\tableline
 \multicolumn{6}{c}{Not-detected in \ci}\\
 \tableline
NGC 1377 & 4.8 & 30 & -- & 0.90 & 0.35\\
NGC 2976 & 0.69 & 27 & -- & 0.10 & 0.10\\
NGC 3077 & 0.74 & 28 & -- & 0.65 & 0.15\\
NGC 5457 & 1.3 & 22 & -- & 0.05 & 0.05\\
 \tableline
 \multicolumn{6}{c}{Central \ci}\\
  \tableline
NGC 1266 & 5.9 & 28 &  -- & 0.95 & 0.45\\
NGC 1482 & 4.4 & 26 & -- & 0.90 & 0.25\\ 
NGC 2798 & 5.0 & 27 & -- & 0.90 & 0.25\\
NGC 3351 & 1.8 & 25 & -- & 0.70 & 0.10\\
NGC 4254 & 2.8 & 22 & -- & 0.20 & 0.10\\
NGC 4536 & 2.8 & 27 & -- & 0.75 & 0.10\\
NGC 5713 & 4.1 & 25 & -- & 0.70 & 0.25\\
 \tableline
\multicolumn{6}{c}{Resolved \ci}\\
 \tableline
NGC 1097 & 2.8 & 25 & 25 & 0.75 & 0.15 \\
NGC 3521 & 2.2 & 21 & 21 & 0.60 & 0.20 \\
NGC 3627 & 1.8 & 24 & 23 & 0.70 & 0.25 \\
NGC 4321 & 2.8 & 23 & 23 & 0.50 & 0.25\\
NGC 4569 & 1.9 & 23 & 22 & 0.50 & 0.10\\
NGC 4631 & 1.5 & 25 & 25 & 0.65 & 0.15\\
NGC 4736 & 0.90 & 27 & 26 & 0.95 & 0.20\\
NGC 4826 & 1.0 & 24 & 23 & 0.95 & 0.20\\
NGC 5055 & 1.5 & 22 & 21 & 0.30 & 0.10\\
NGC 6946 & 1.3 & 26 & 23 & 0.50 & 0.30\\
NGC 7331 & 2.8 & 22 & 22 & 0.75 & 0.20\\

\tableline
\end{tabular}
\end{center}
\tablecomments{Modified blackbody derived temperatures for the centers of the BtP galaxies and the average dust temperatures of regions detected in \ci for their disks. The proportion of 70\micron emission enclosed and the fraction of the $R_{25}$ galactic radius covered by the regions detected in \ci (or the central aperture for non-detections) are shown in the last two columns (rounded to nearest 5\% to indicate rough level of precision).}
\end{table}

\startlongtable
\begin{deluxetable*}{lccccccc}
\tablecaption{\ci line intensities\label{tab:cidata}}
\tablehead{
\colhead{Galaxy} & \colhead{Reg.} & \colhead{RA} & \colhead{Dec.} & \colhead{$I([\mathrm{CI}]_{(1-0)})$} & \colhead{Err($I([\mathrm{CI}]_{(1-0)})$)} & \colhead{$I([\mathrm{CI}]_{(2-1)})$} & \colhead{Err($I([\mathrm{CI}]_{(2-1)})$)} }
\startdata
 &  & Deg. & Deg. & $\mathrm{W\,sr^{-1}\,m^{-2}}$ & $\mathrm{W\,sr^{-1}\,m^{-2}}$ & $\mathrm{W\,sr^{-1}\,m^{-2}}$ & $\mathrm{W\,sr^{-1}\,m^{-2}}$ \\
\hline
NGC1097 & 1\_1 & 41.58504 & -30.26998 & 5.178E-10 & 9.383E-11 & 6.378E-10 & 6.313E-11 \\
NGC1097 & 2\_1 & 41.57861 & -30.27002 & 7.368E-10 & 1.037E-10 & 9.7E-10 & 5.456E-11 \\
NGC1097 & 3\_1 & 41.57218 & -30.27006 & 3.98E-10 & 6.825E-11 & 5.191E-10 & 4.634E-11 \\
NGC1097 & 0\_2 & 41.59152 & -30.27549 & 1.759E-10 & 1.203E-10 & 2.856E-10 & 7.672E-11 \\
NGC1097 & 1\_2 & 41.58509 & -30.27553 & 7.571E-10 & 9.933E-11 & 1.055E-09 & 5.636E-11 \\
\enddata
\tablecomments{The second column indicates the region name for the Beyond the Peak survey and will allow easy cross-referencing to the CO line intensities. The regions are enumerated as `x\_y' in a coordinate grid, with `2\_2' the central region. Table \ref{tab:cidata} is published in its entirety in the machine-readable format.
      A portion is shown here for guidance regarding its form and content.}
\end{deluxetable*}

\subsection{Auxiliary data}

Ground-based CO data were compiled from multiple sources. CO(1-0) data came from the BIMA-SONG survey \citep[only those galaxies with short-spacings;][]{helfer03} or the Nobeyama CO Atlas of Nearby Spiral Galaxies \citep{kuno07}. CO(2-1) data came from the HERACLES survey \citep{leroy13} and CO(3-2) data came from the JCMT Nearby Galaxies Legacy  Survey and archival data \citep{wilson12}. The datacubes from these surveys were convolved to match our spatial resolution; line fluxes were extracted based upon velocity ranges from the high signal-to-noise CO(2-1) data.  $\alpha_{\mathrm{CO}}$ values are from \citet{sandstrom13}, where available. 

\subsection{Modified blackbody dust temperatures}

\begin{figure}
\begin{center}
\includegraphics[width=8cm]{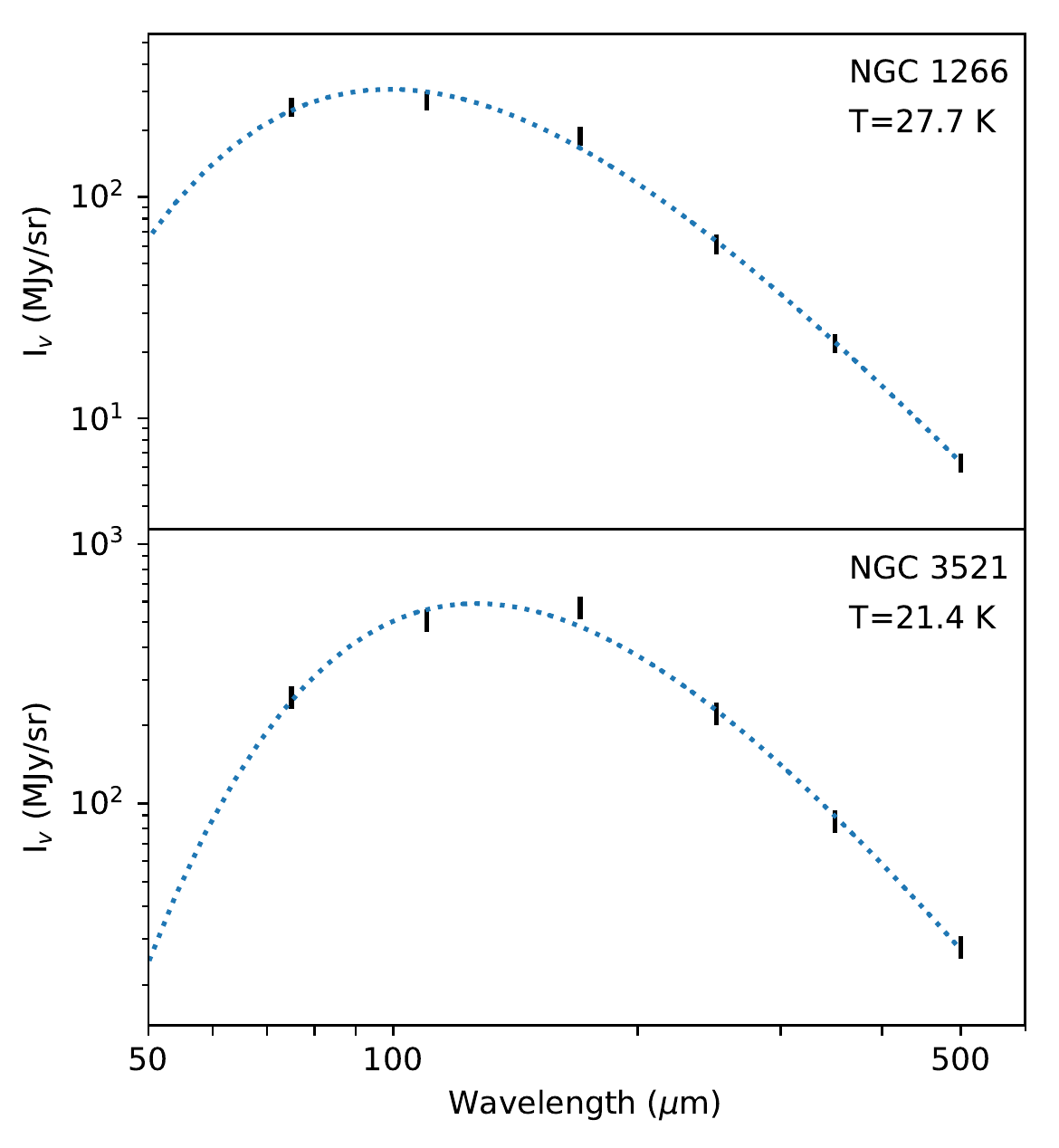}

\end{center}
\caption{ Two examples of the single component modified blackbody fits to the FIR spectral energy distribution from 70 to 500 $\mu$m. }
\label{fig:dustfit}
\end{figure}

We fit single-component modified blackbody (MBB) functions to the FIR photometry for regions matching our spectroscopic apertures:

\begin{equation}
I_{\nu} = A  \nu^{\beta} B_{\nu}(T),
\end{equation} where $A$ is an amplitude, $\beta$ is the emissivity, and $B_{\nu}(T)$ is the Planck function at temperature $T$. We use far-IR photometry at 70, 160, 250, 350 and 500 \micron from the Herschel PACS and SPIRE maps obtained under the KINGFISH project \citep{kennicutt11}. The fits were performed using curve\_fit function from the SciPy library; two example fits are shown in Fig.~\ref{fig:dustfit}. The fit parameters give an estimate of the cold dust temperature and the $\beta$ parameter modifying the blackbody shape. We choose to use a single MBB component and drop the 24 \micron photometry available from Spitzer,  because the addition of this one more point, but three more free parameters (associated with a second MBB component) led to more degeneracies.

\section{Empirical Analysis}

We start with an empirical analysis of the spatially resolved \ci and CO line data. This approach has the inherent strength of being independent of any model assumptions. Critical densities and upper level temperatures for the lines involved are listed in Table~\ref{tab:linepars}.

\begin{table}[htp]
\begin{center}
\caption{Reference line parameters.}
\label{tab:linepars}
\begin{tabular}{lccc}
\tableline
Line & $T_{\mathrm{u}}$ (K) & $n_{\mathrm{crit}}$ (cm$^{-3}$)&  $n_{\mathrm{crit}}$ (cm$^{-3}$)\\
 & & 10 K & 100 K \\
\tableline
\cione & 24 & $1.1 \times 10^{3}$ & $1.1 \times 10^{3}$ \\
\citwo & 62 & $1.8 \times 10^{3}$  & $1.2 \times 10^{3}$ \\
CO(1-0) & 5.5 & $1.9 \times 10^{3}$ & $2.1 \times 10^{3}$ \\
CO(2-1) & 17 & $6.1 \times 10^{3}$ & $6.7 \times 10^{3}$\\
CO(3-2) & 33 & $1.5 \times 10^{4}$ & $1.7 \times 10^{4}$\\
CO(4-3) & 55 & $3.3 \times 10^{4}$ & $3.4 \times 10^{4}$\\
CO(5-4) & 83 & $6.3 \times 10^{4}$ & $6.1 \times 10^{4}$\\
CO(6-5) & 116 & $1.1 \times 10^{5}$ & $9.8 \times 10^{4}$ \\
CO(7-6) & 155 & $1.5\times 10^{5}$ & $1.5 \times 10^{5}$\\
\tableline
\end{tabular}
\end{center}
\tablecomments{Critical densities and upper level temperatures for relevant lines. Critical densities are calculated assuming ortho-H$_{2}$ is the dominant collision partner and using rate coefficients appropriate for a kinetic temperature of either 10K or 100K.}
\end{table}

\subsection{Disk versus central emission}

\begin{figure*}
\begin{center}
\includegraphics[width=16cm]{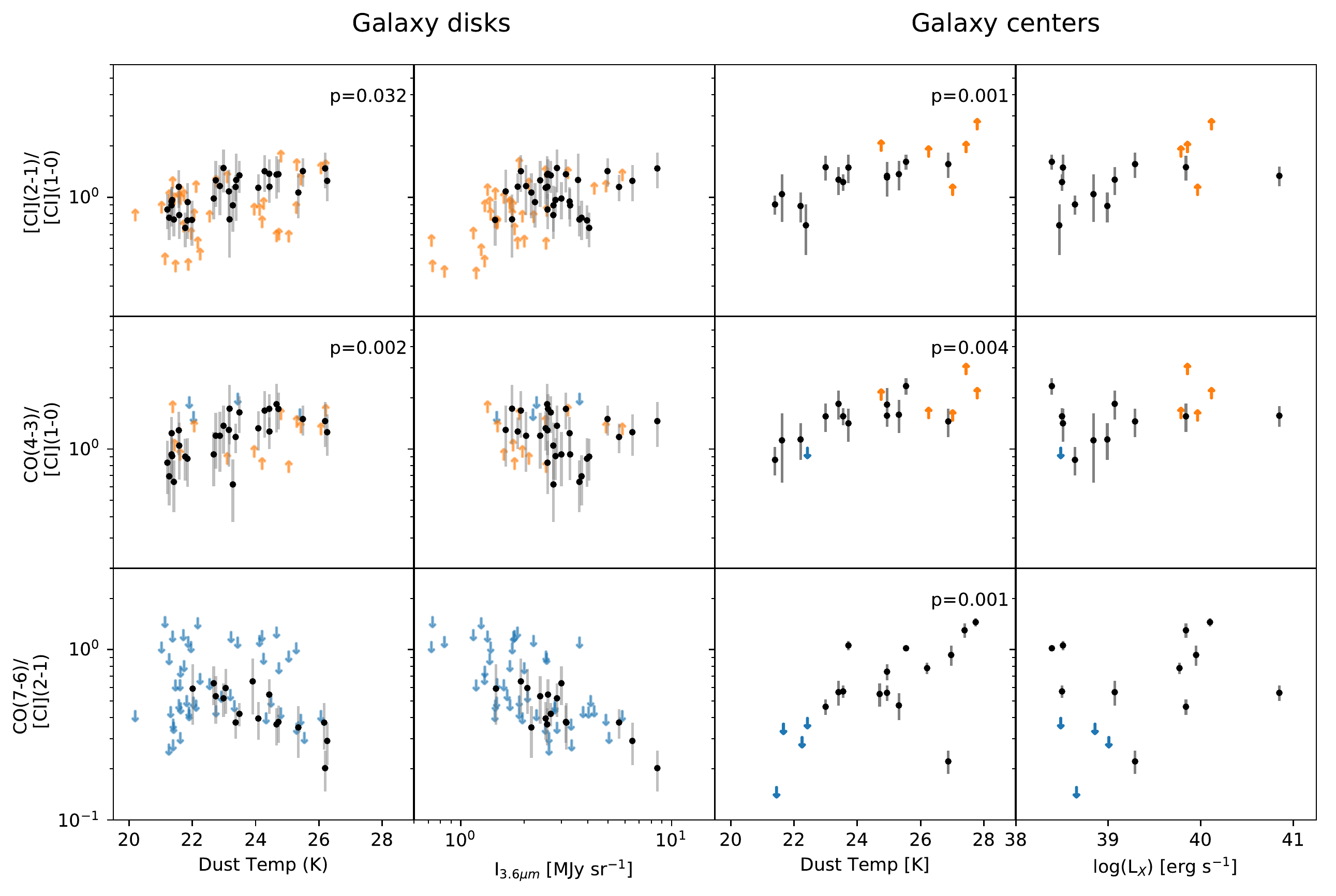}

\end{center}
\caption{ \citwo/\cione, CO(4-3)/\cione and CO(7-6)/\citwo ratios for galaxy disks and centers as a function of dust temperature (1st and 3rd columns), IRAC 3.6 \micron intensity (2nd column, disks only) and central X-ray luminosity (4th column, centers only). Upper (lower) limits are shown as blue (orange) arrows. }
\label{fig:gal_dc}
\end{figure*}

We separate the disk emission from the central apertures in order to investigate the properties of \ci \xspace emission in normal galaxy disks as well as in isolated galaxy centers. We plot the disk and central \citwo/\cione, CO(4-3)/\cione and CO(7-6)/\citwo \xspace ratios against dust temperature, IRAC 3.6 \micron intensity (for the disks; a proxy for stellar surface density) and central X-ray luminosity (for the centers) in  Fig.~\ref{fig:gal_dc}.  All three ratios are chosen such that higher values reflect higher excitation conditions (temperature or density, see Table~\ref{tab:linepars}). The motivation for pairing these specific CO lines with the \ci \xspace lines is that they are the closest in frequency and thus may be conveniently observed together, with similar angular resolutions. With 40\arcsec~diameter apertures separated by 40\arcsec~between aperture centers, we consider all non-central apertures to be dominated by disk emission. 
 
We look for correlations between both the disk and central \ci \xspace ratios with dust temperature, as a proxy of local ISM excitation. Using a Kendall's tau statistic appropriate for censored data as implemented in the R package NADA \citep[Nondetects And Data Analysis;][]{lee17}, we can reject the null hypothesis of no correlation for the \citwo/\cione and CO(4-3)/\cione \xspace ratios for both disks and galaxy centers and the CO(7-6)/\citwo \xspace ratio for galaxy centers. The $p$-values of these correlations are shown in the upper right-hand corner of the appropriate sub-plot in Fig.~\ref{fig:gal_dc}. In all of these cases, the data show the expected correlation of more highly excited \ci \xspace ratios with increasing dust temperature.

There are very few detections of the CO(7-6)/\citwo \xspace ratio within galaxy disks, but surprisingly, the detections tend to decrease in ratio as dust temperature increases in contrast to what is seen for the galaxy centers and reported elsewhere for LIRGs \citep{lu17}. We note that all of the high-temperature low CO(7-6)/\citwo \xspace ratio points are contributed by the same galaxy, NGC~4736. This galaxy's central near-IR spectrum lacks Br$\gamma$ emission \citep{walker88} and optical spectroscopy diagnoses its nuclear region as a 1~Gyr post-starburst with very little current star formation \citep{taniguchi96}. NGC 4736's dust is likely heated by this high density poststarburst population instead of ongoing star formation. Furthermore, NGC~4736 has the strongest central \cii suppression \citep{smith17} which is similarly interpreted to be due to a high-intensity, but softer (than ongoing star formation) stellar radiation field. Thus NGC~4736's very low CO(7-6)/\citwo ratios at relatively high dust temperature may be explained by intense starlight with a relative lack of recent star formation.
 
Within galaxy disks, the IRAC 3.6 \micron intensity is a proxy for the density of the radiation field contributed by old stars which has been found to correlate with the \cii deficit \citep{smith17}. The relations in Fig.~\ref{fig:gal_dc} do not show any statistically significant correlations between these three \ci  ratios and the IRAC 3.6 \micron intensity. Note that galaxy centers are excluded from this plot, because a low-luminosity AGN or central starburst can make even the 3.6 \micron emission a less reliable tracer of old stars.

The galactic centers of the BtP galaxies are a mix of low-luminosity AGN, nuclear starbursts and quiescent nuclei \citep{moustakas10} in otherwise normal galaxies. Due to the resolution of our observations, the central extraction measures not only the true nucleus, but extended circumnuclear regions from 0.9 to 6 kpc scale (see Table~\ref{tab:sample}). Line ratios for these central regions are plotted against their nuclear (within 2\farcs3) X-ray luminosity  in Fig.~\ref{fig:gal_dc} in order to look for correlations with AGN or star formation power. The majority of these X-ray luminosities are from the compilation by \citet{grier11} of Chandra measurements, with additional Chandra measurements for NGC~1266 \citep{alatalo14} and NGC~4536 \citep{mcalpine11}. Note these X-ray luminosities include both hard and soft emission (0.3-8.0 keV) and so may have contributions from an AGN and/or from central star-formation. No statistically significant  correlations are observed between X-ray luminosity and any of these central \ci ratios.

\subsection{ \ci and CO}

\begin{figure*}
\begin{center}
\includegraphics[width=17cm]{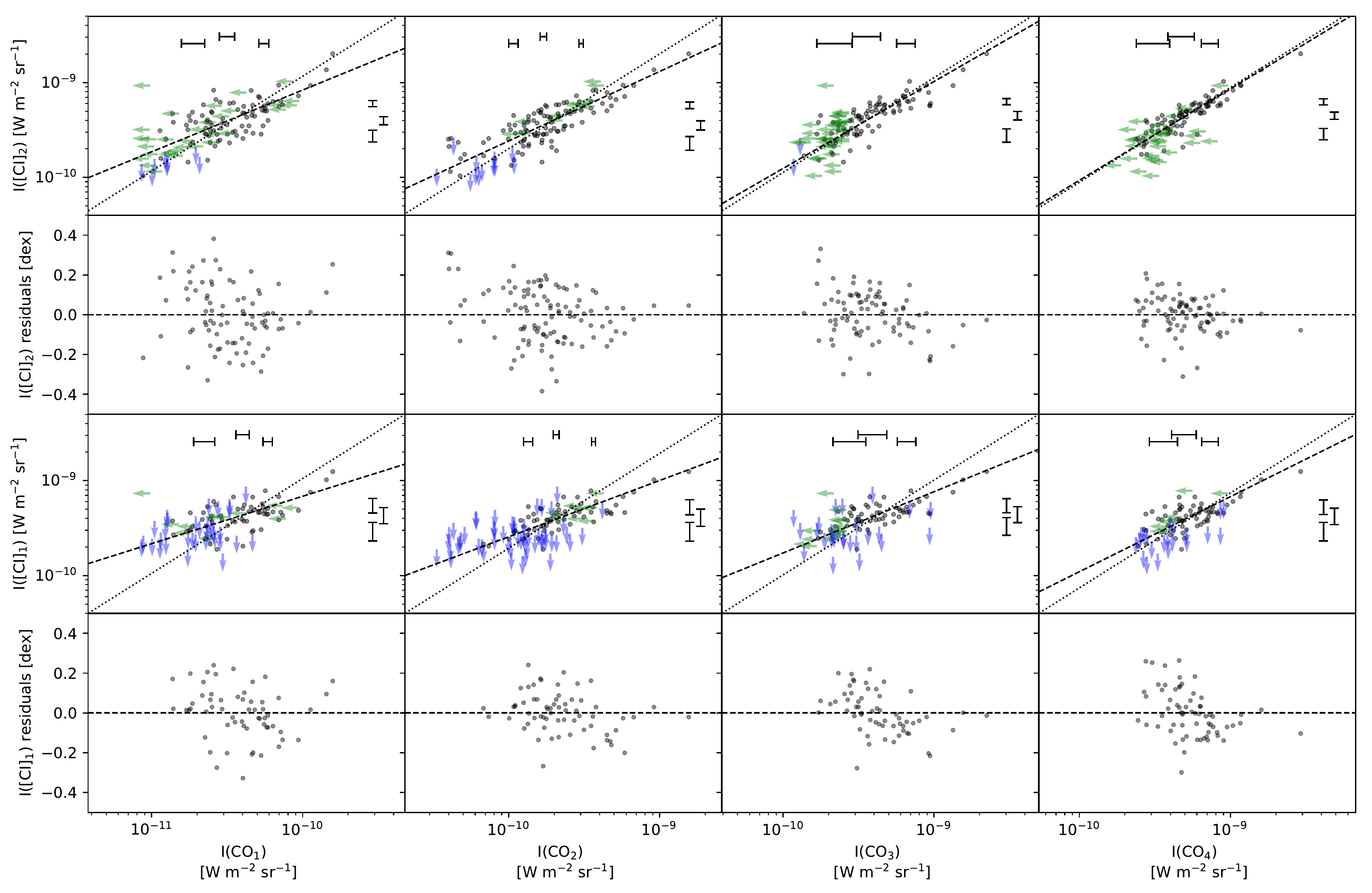}
\includegraphics[width=17cm]{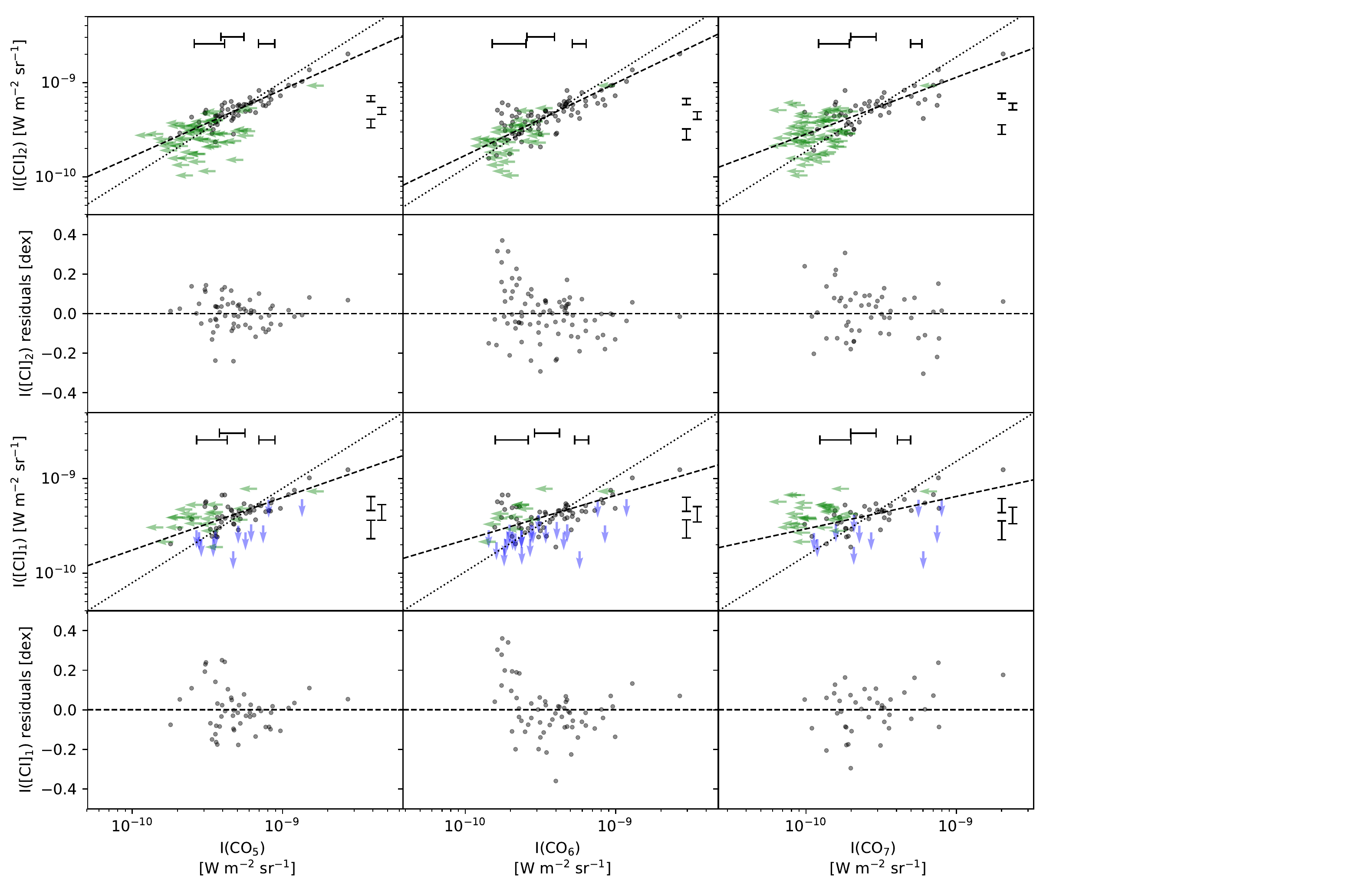}

\end{center}
\caption{ \citwo (first two rows) and \cione (second two rows) versus all seven CO lines. The first and third rows of each column are intensity versus intensity plots while the second and fourth rows show logarithmic residuals from the best-fit power law (dashed line). The dotted line in the intensity versus intensity plots indicates a relation with a linear slope. Floating errorbars above and to the right depict typical errors for the BtP or ground-based detected data in three intensity bins for each ordinate.}
\label{fig:ci_vs_co}
\end{figure*}

Both [CI] lines are indisputably correlated with all observed transitions of CO, as shown in the intensity versus intensity plots of  Fig.~\ref{fig:ci_vs_co} (first and third rows of each column).  In these plots, grey circles indicate detections, green left arrows indicate upper limits in CO and blue down arrows indicate upper limits in \ci. For clarity, data which are non-detections in both lines are not shown. We fit a power law relation between the intensities (indicated with a dashed line) and the residual scatter around this relation (second and fourth row of each column). 

The power-law we are interested in is written:
 \begin{equation}
 I(\mathrm{[CI]}) \propto I(\mathrm{CO})^\gamma.
 \end{equation}
Values of $\gamma$ close to one imply that it is reasonable to convert directly from one line intensity to the other. Sub- or super-linear correlations imply a dependence on a physical effect that correlates with the line intensity of a region. Fitting in log-log space means the slope of a linear fit will be the desired power-law exponent, $\gamma$. However, the task of fitting such a line is complicated by the multitude of upper limits within these datasets. We follow advice from \citet{feigelson12} and use the doubly-censored Theil-Sen estimator given in \citet{akritas95} for fitting lines with doubly left-censored data as implemented in the R package NADA.  We determine best slopes using this algorithm for each combination of CO line and \ci line, which are shown with dashed lines in Fig.~\ref{fig:ci_vs_co}. It should be noted that this algorithm does not return a best estimate for the y-intercept. Thus we used least squares to find a y-intercept estimate having fixed the slope at the NADA value and ignoring the censored data.

\begin{figure}
\begin{center}
\includegraphics[width=8.6cm]{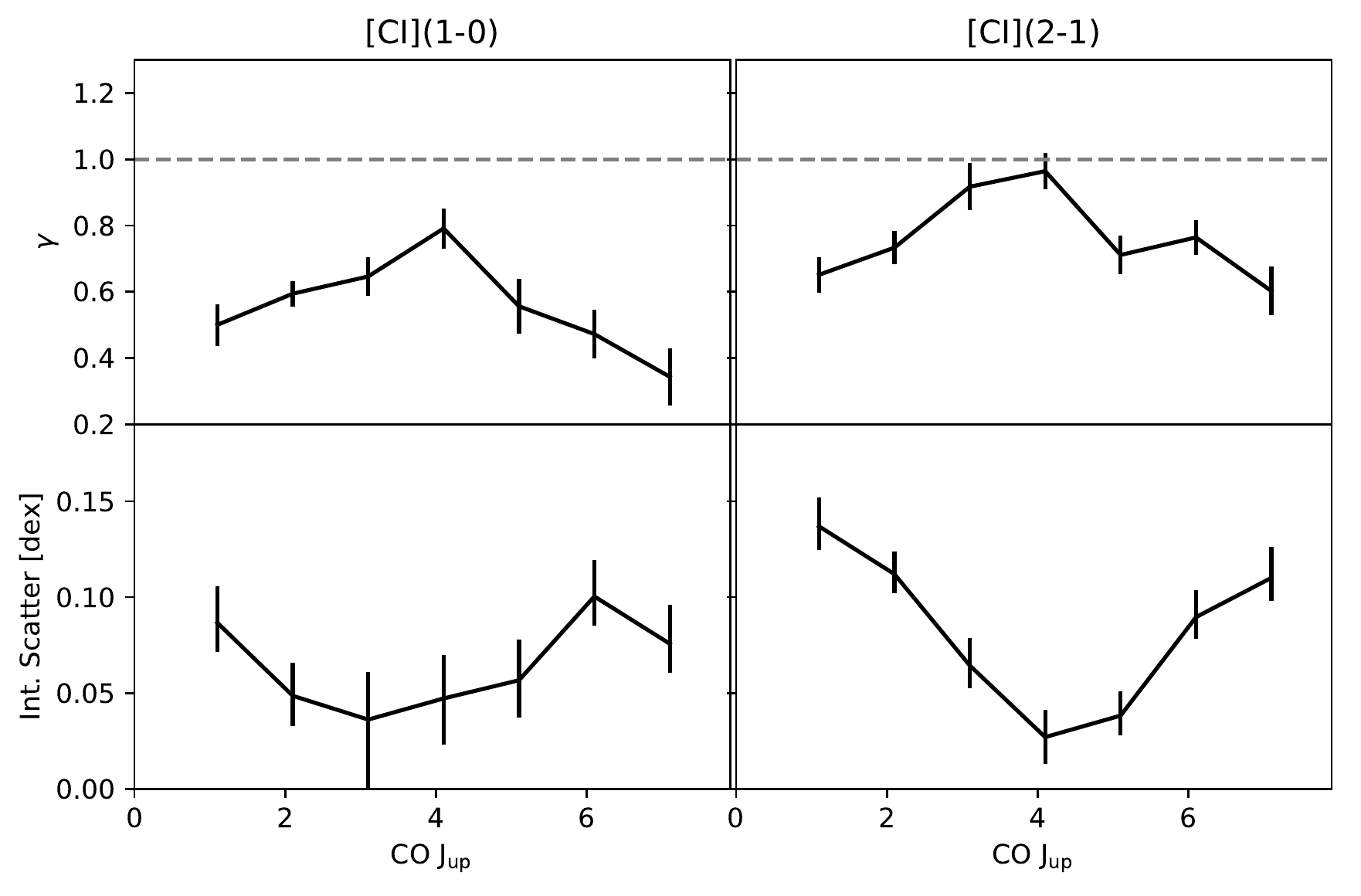}
\end{center}
\caption{Power law exponent, $\gamma$, (top) and intrinsic scatter (bottom) as calculated by a power law fit of \cione and \citwo intensities versus CO line intensities of Fig.~\ref{fig:ci_vs_co}. The value the CO $J_{\mathrm{up}}$ is shown along the x-axis.}
\label{fig:lts_both}
\end{figure}

The calculated values of the power-law exponent are shown in Fig.~\ref{fig:lts_both} for both \cione and \citwo against the various $J$ CO lines.  Uncertainties on these slopes are determined by bootstrap resampling of the data. For \citwo, the relationship with CO(4-3) is very nearly linear. For \cione, the relationship between \ci and CO is always sublinear, so this means
that higher intensity CO regions have higher CO/\ci ratios, with a larger effect for lines further from CO(4-3). For both \cione and \citwo, we expect the following physical reasons are at work. 
At high intensities of low-J CO, the majority of gas is cold, and unable to excite the transitions of  \ci \xspace with their higher upper level temperatures. Thus these regions have lower \ci/CO ratios and result in sublinear correlations. At high intensities of high-$J$ CO, there is more dense, warm gas which is more effective at exciting the high critical density high-$J$ CO lines. Again this results in lower \ci/CO ratios at the high intensity end and thus, again, a sublinear correlation. 

The intrinsic scatter in the relation can be determined using the statistical approach of  \citet{cappellari13} (their Eqn.~6). The basic idea is to compute the reduced $\chi^2$ statistic including a term for the intrinsic scatter which may be increased until the reduced $\chi^2$ equals its expected value of 1. Using this approach, the intrinsic scatter appears minimal for the \citwo to CO(4-3) relation as shown in the bottom panels of Fig.~\ref{fig:lts_both}. This low scatter is a further indication of a strong relationship between \citwo and CO(4-3). Relations between \cione and CO(2-1), CO(3-2) and CO(4-3) all have similar intrinsic scatter. Thus, we cannot make a strong statement about which CO line \cione is best correlated with.

\section{An approximation and Models}

\subsection{LTE approximation}

\begin{figure}
\begin{center}
\includegraphics[width=8cm]{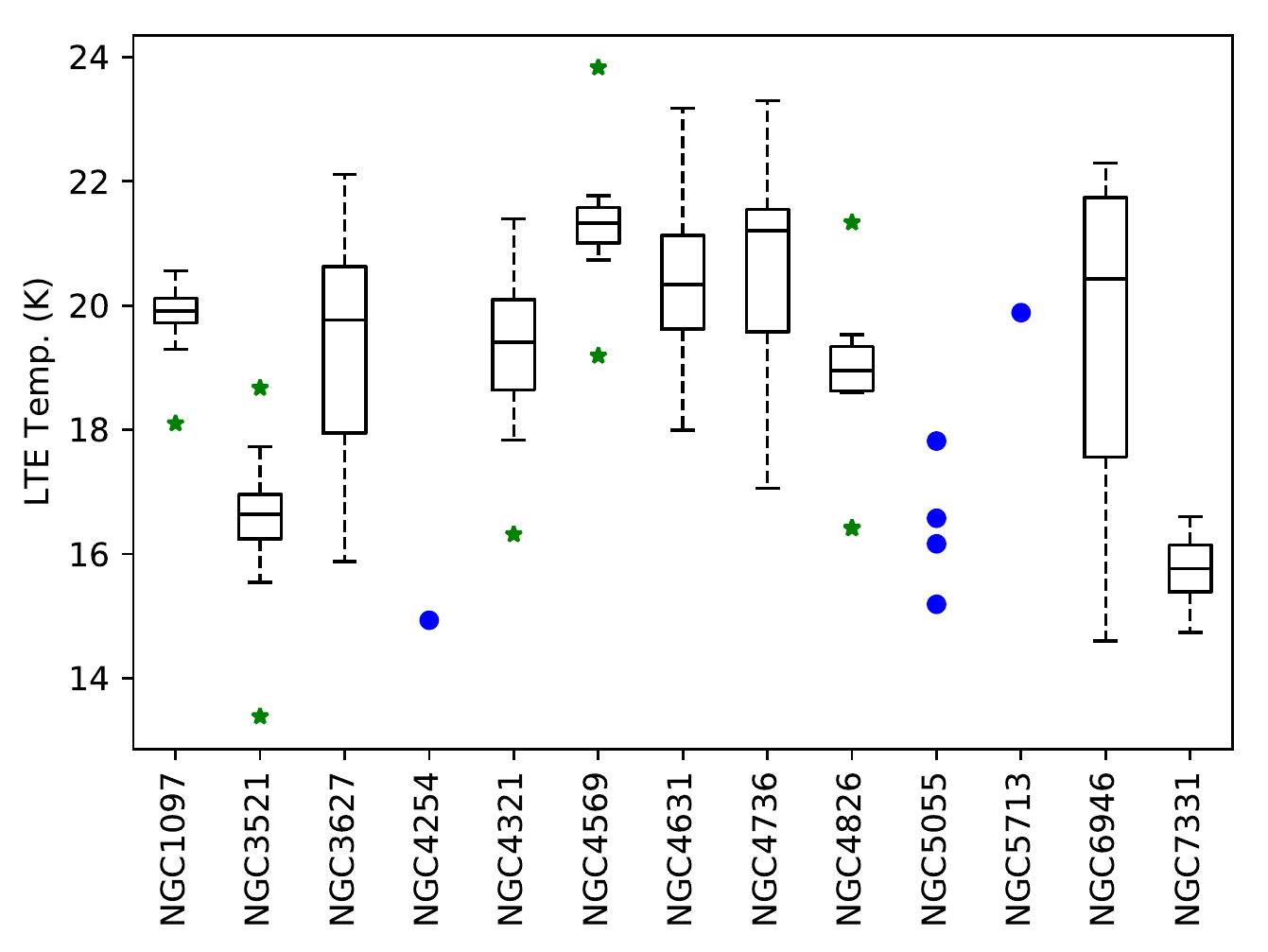}

\end{center}
\caption{Boxplots indicating the distribution of LTE \ci temperatures for each galaxy. For galaxies with fewer than six \ci temperature measurements, blue dots show the individual temperatures instead of a boxplot. Green stars represent outliers in the distributions.}
\label{fig:CI_temp}
\end{figure}

We first consider what the \ci lines alone can tell us about the physical state of the gas. Unfortunately, there are only two \ci lines, which means we must make some assumptions if we are to derive gas properties. A common assumption is that the neutral carbon is in LTE and the two \ci lines are optically thin \citep[e.g.][]{kamen2014}. In this case, the excitation temperature is simply calculated by using the Boltzmann distribution for the \ci level populations and Einstein A coefficients to determine the line fluxes from these level populations. Fig.~\ref{fig:CI_temp} documents these optically-thin LTE temperatures for all galaxies where both \cione and \citwo are detected. These temperatures are generally lower than those found for the more active galaxies (starbursts, AGN, ULIRGs) in the \citet{kamen2014} sample. Within their sample, only six out of eighteen studied regions have LTE \ci temperatures below 24~K, the highest temperature seen in the BtP sample. These optically-thin LTE temperatures are consistent with the kinetic temperatures found for the \ci-emitting gas for simulated molecular clouds at both a solar neighborhood interstellar radiation field and an ISRF ten times stronger \citep{clark19}.

 However, the conditions for LTE may not be met, for instance, if subthermal excitation is important. Section 3.3 of \citet{glover15} discusses the problems they see when assuming LTE for the \ci emission of their simulated cloud. If lines are (partially) optically thick, we also cannot as simply derive the excitation temperature, as the observed flux will be lower than the emitted flux. We will later discuss how important these effects are for the BtP sample, if we use the gas properties derived from our two-component fit.

\subsection{Two-component LVG + template models}

Next we use models to help connect the \ci lines to a physical gas phase. As CO and  \ci \xspace emission have been observed to be cospatial on parsec scales within Milky Way clouds (e.~g. Figures 3 and 4 of \citet{shimajiri13}) as well as on hundreds of parsec scales within resolved galaxies (e.~g. \citet{krips16}), we use the emission from both species to constrain the CO- and \ci-emitting gas parameters.

Large velocity gradient (LVG) models are commonly used to describe the emission from molecular clouds because they provide an easily computable non-LTE approximation. To model the line intensities of a single molecular species, the parameters required are the kinetic temperature, the volume density of dominant collision partners and the column density of the molecule of interest per unit velocity spread. When such models are applied to regions within galaxies, we assume those regions are made up of clouds with identical values of these parameters. An additional normalization parameter allows for the intensity of the region to be lower than if it were uniformly filled by such clouds. LVG models may also include multiple species, so long as either an abundance ratio is assumed, or there are enough lines to allow the abundance ratio to be fit. 

Two-component LVG models are required to fit the \ci and CO emission of these galaxies. \citet{israel15} note that their lowest-excitation galaxies cannot be well fit by a single component. \citet{schirm17} document in their Appendix A why a one-component LVG model is not realistic for regions in M51. Similarly, we found unrealistically high gas kinetic temperatures and low densities when the BtP data were fit with a single LVG component.

A full two-component model requires ten free parameters: density, kinetic temperature, column density per velocity interval, filling factor and C to CO abundance for both components. In the very best cases within the BtP sample, only 9 lines were detected, and therefore a more constrained model was necessary. We first tried a combination of a low-excitation template based upon the Milky Way and fit a high-excitation LVG model. However, this led to high reduced $\chi^2$ values for many regions due to poor fits at the low-excitation end of the spectral line energy distribution (SLED). Thus, we instead use an empirically-derived template for the high-excitation emission and an LVG model to account for the lower-excitation emission. The high-excitation template we use is based on the detailed empirical observations of an actively star-forming region in the LMC, N159W \citep{lee16}. \citet{lee16} analyze the sources powering the \ci and CO emission in this region, acknowledging some is from star-formation powered PDRs, while other emission must be powered by star-formation feedback related shocks. Instead of modeling these components separately, we simply use their integrated observed emission from the region as a template. 
For the lower-excitation component, we use the LVG code RADEX to provide models of the emission \citep{vandertak07}. We constrained the models to $n(\mathrm{H}_{2}) = 10^{3}$ cm$^{-3}$ because at higher densities the shape simply resembles the high-excitation template and we lose diagnostic power. Kinetic temperature is allowed to vary from 10 to 35 K in 5 K increments.  $N(\mathrm{CO})/\Delta v = 0.25, 0.5, 1, 2, 4 \times 10^{17}$ cm$^{-2}$~km$^{-1}$~s are options for the column density per unit velocity. In addition, we allow the C$^{0}$:CO abundance ratio to be 2:1, 1:1, 1:2, 1:4, 1:8 or 1:16. These choices are all based upon the `quiescent' and `very quiescent' models for \ci and CO from \citet{papadopoulos04}, which summarize observations of the Milky Way, M31 and M33. They also cover the parameters found to fit the cold component of M51 \citep{schirm17}. The scale factor for each component is defined to be the average number of $\Delta v = 1$ km s$^{-1}$ clouds along the line of sight and is hereafter denoted $f^{\star}_\mathcal{L}$. While this appears to be an awkward definition, it prevents us from either having to fix the typical cloud width within our galaxies or from assuming the full velocity line width of the galaxy is the relevant line width for the LVG approximation. Based upon this definition, the N159W high-excitation region has $f^{\star}_\mathcal{H} = 1$ because it is a 10 km s$^{-1}$ wide cloud with a beam filling factor of 0.1 \citep{lee16}. Thus the scale factor we apply to the high-excitation N159W template for a given region gives us that region's $f^{\star}_\mathcal{H}$ scale factor.

\begin{figure}
\begin{center}
\includegraphics[width=8cm]{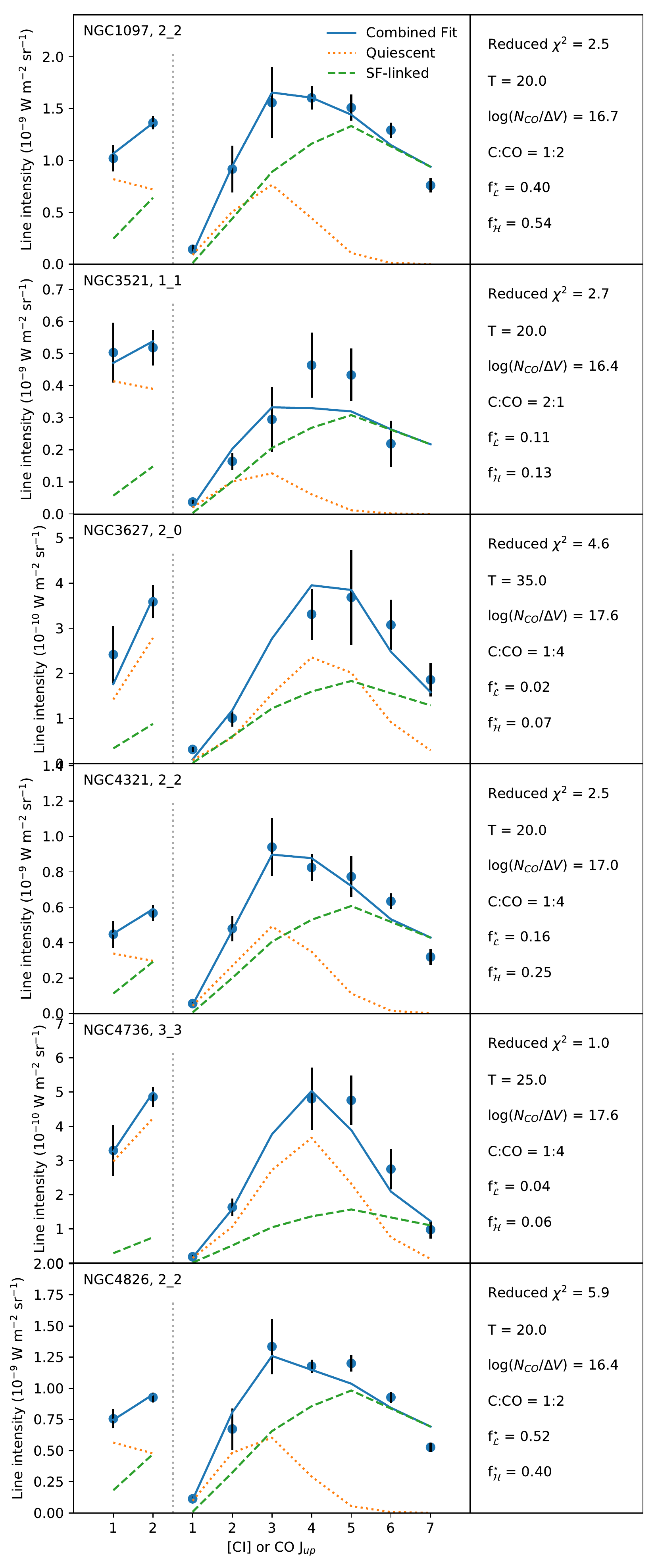}
\end{center}
\caption{Six example fits to \ci and CO SLEDS. The fits combine a constant shape high-excitation template (green dashed line) and a low-excitation component (red dotted line) based on RADEX models with varying temperature, column density per velocity unit and C to CO abundance ratio. Both the high-excitation template and low-excitation component are scaled by a filling factor, $f^{\star}_{\mathcal{H}}$ and $f^{\star}_{\mathcal{L}}$, respectively.  }
\label{fig:sleds}
\end{figure}

\begin{figure}
\begin{center}
\includegraphics[width=8cm]{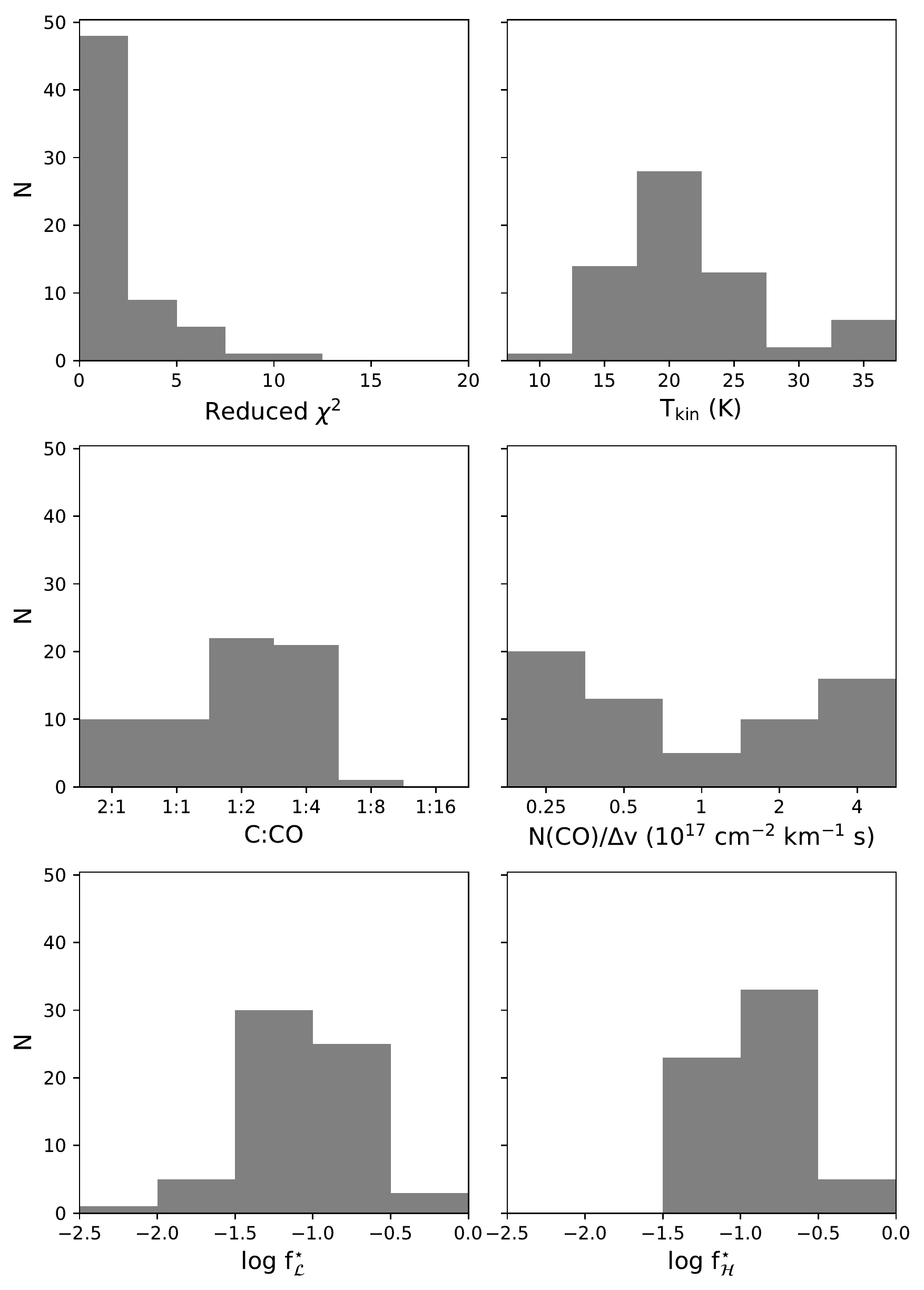}
\end{center}
\caption{Histograms representing the results of the two-component fit to the \ci and CO lines. Best-fit models for different regions vary over the full range of $T_{\mathrm{kin}}$ and $N(\mathrm{CO})/\Delta v$, although few regions are best fit by 10 K. Very low C:CO abundances do not seem to be required. Filling factors are not remarkably different between low and high excitation components. Reduced $\chi^{2}$ values are mostly lower than 5.}
\label{fig:result_histos}
\end{figure}

For robust fits of the five parameters ($T$, $N(\mathrm{CO})/\Delta v$, C:CO, $f^{\star}_\mathcal{L}$, $f^{\star}_\mathcal{H}$)  we decided to require detections in both \ci lines and at least four CO lines, including one of CO(1-0) or CO(2-1). This cut results in 64 independent regions that are fit. The best fit is determined by minimizing the $\chi^2$. We show the data and best fits for six example regions in Fig.~\ref{fig:sleds}; histograms documenting the fit results for all regions are shown in Fig.~\ref{fig:result_histos}.

\subsection{Excitation fraction distributions}

\begin{figure}
\begin{center}
\includegraphics[width=8cm]{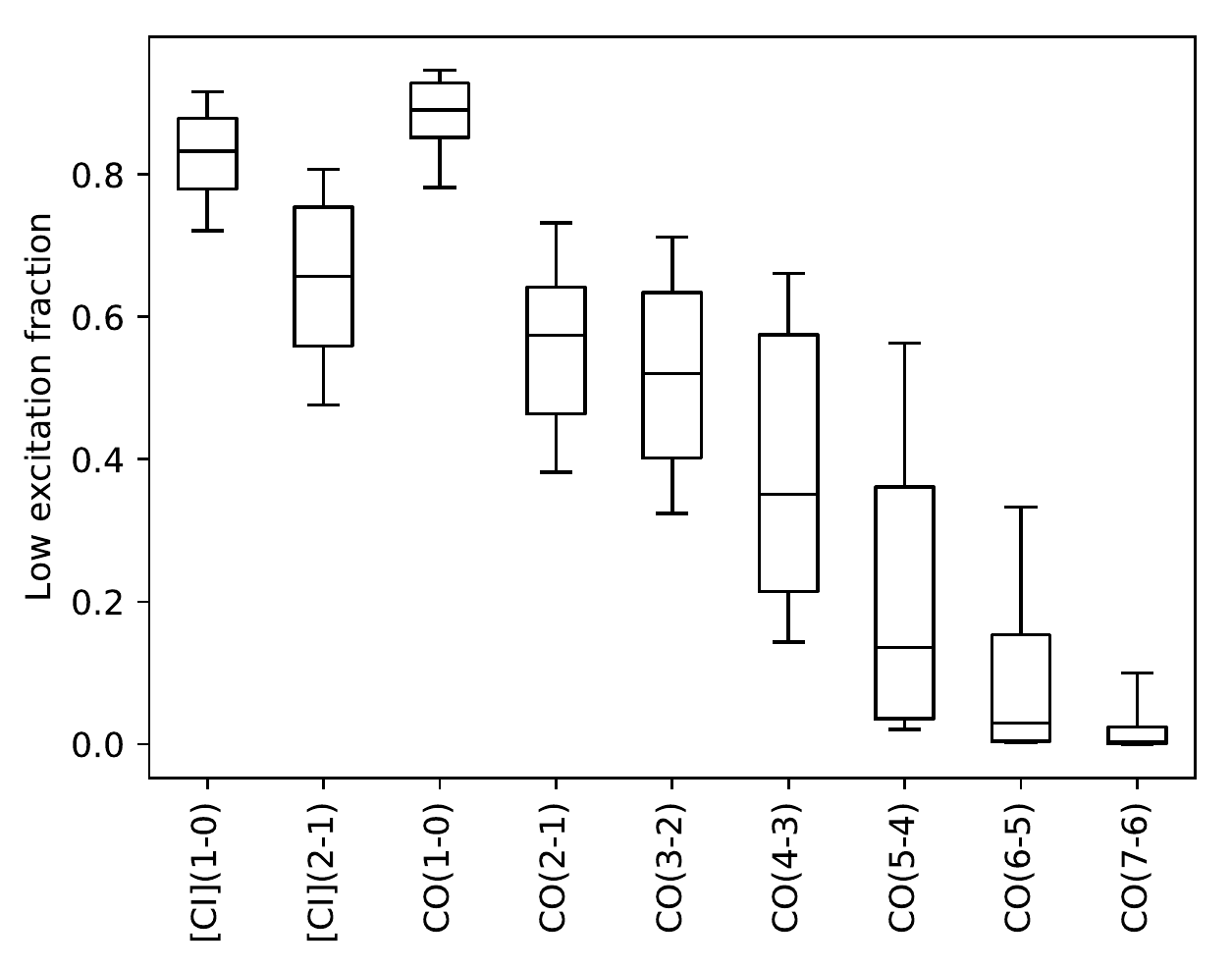}
\end{center}
\caption{Distributions of the fraction of line flux contributed by the low-excitation component, according to the two-component model fit. The boxes enclose the middle two quartiles, with the median indicated. The whiskers extend to 10\% and 90\% of the full data range.}
\label{fig:cold_frac}
\end{figure}

Given the adequate decomposition provided by the two-component fit, we indicate distributions of the low-excitation fraction of each line in Fig.~\ref{fig:cold_frac}. For \ci, nearly all of the \cione line is contributed by the low-excitation component. For \citwo, the fraction varies significantly from region to region although the median is close to 70\%.  For CO, there is a steady decrease in the contribution of the low-excitation component as $J$ increases.  We note that our range for the CO(2-1) high-excitation fraction corresponds nearly identically to the range \citet{israel15} report for their LIRG/ULIRGs CO(2-1) dense fraction (rightmost panel of their Fig. 8). As expected, very little of the highest-J CO line intensities are contributed by the low-excitation component.

These distributions represent the fraction of line emission from a low-excitation component for the brightest regions within a sample of normal, local spiral galaxies. By this analysis, \cione and CO(1-0) are both reliable tracers of the low-excitation component for these regions. \citwo is less reliable as even in these normal spiral galaxies there are cases where half of its emission is from a high-excitation component. The CO $J_{\mathrm{up}}=2-5$ lines have low-excitation fraction ranges of typically 30-70\%. So for these regions of spiral galaxies they are not dominated by one component or the other and indeed show a wide variation. While this limited two-component analysis is only a slight improvement on trying to account for the physical properties with a single LVG model, already we can see that considering multiple components is very important for the \ci and CO emitting gas. 

\subsection{The \citwo to CO(4-3) connection}

\begin{figure}
\begin{center}
\includegraphics[width=8.5cm]{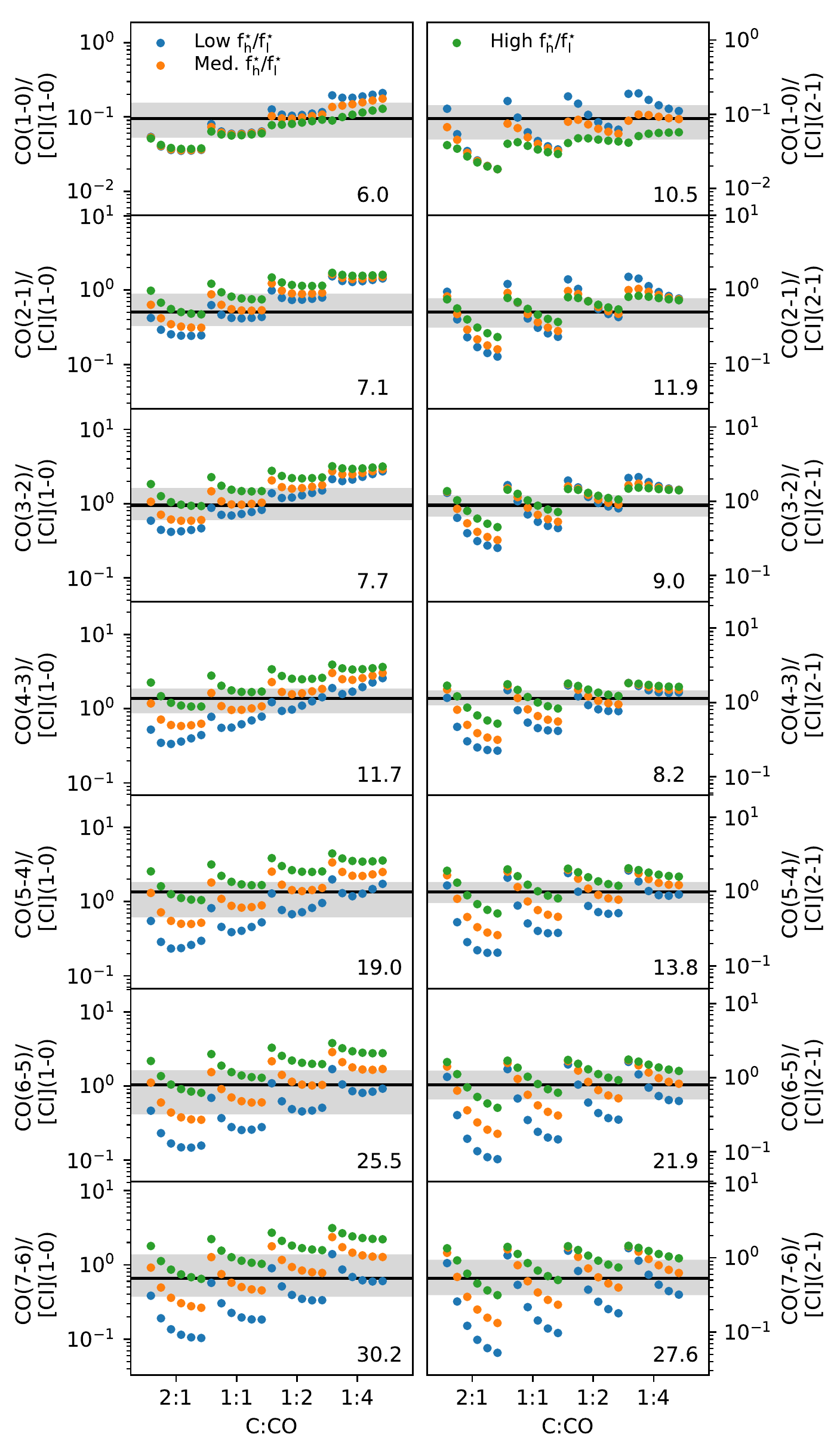}
\end{center}
\caption{Variation in model ratios with C:CO abundance, temperature, and filling factor ratio with a fixed $N(\mathrm{CO})/\Delta v = 10^{17}$ cm$^{-2}$ km$^{-1}$ s. The left column presents CO ratios with \cione and the right column with \citwo. Along the x-axis, different models are indicated, first by C:CO abundance of the low-excitation component, then within those categories, by low-excitation component temperature from 10 to 35 K in 5 K increments (thus 6 points per abundance flight). Blue, orange, green points indicate filling factor ratios at 10\%, 50\% and 90\% of the filling factor ratio distribution. The ratios of these models may be compared to the observed ratios for each combination, indicated by the median (black line) and the zone containing 10\% to 90\% of the distribution for the BtP sample, in grey. The number in the bottom right indicates the factor spread from the highest to lowest ratio of the models.}
\label{fig:blah2}
\end{figure}

One of the goals of this modeling work is to explain the strong link found between the CO(4-3) and \citwo emission. In order to do this, we consider the range of line ratios produced by these two-component models. For various abundances and temperatures (shown on the x-axis; temperature increases from 10 to 35 K left to right in 5 K steps for each abundance), Fig.~\ref{fig:blah2} shows the line ratios produced by our two-component model. The different color circles represent different $f^{\star}_\mathcal{H}/f^{\star}_\mathcal{L}$ filling factor ratios. We depict the median ratio in orange, the ratio at 90\% of the observed distribution in green and the ratio at 10\% of the observed distribution in blue. We only show the $N(\mathrm{CO})/\Delta v = 10^{17}$ cm$^{-2}$ km$^{-1}$ s models to avoid clutter. The number in the bottom-right corner shows the multiplicative spread from the minimum to maximum line ratio for the models. 

For \citwo, the models collapse to a small range at the CO(4-3)/\citwo ratio. Note that this also occurs for the four other column densities individually and all five column densities as a set (not shown). This collapse fits precisely with the linear correlation with minimal scatter observed for CO(4-3) versus \citwo in Section~3.2. While Fig.~\ref{fig:blah2} graphically shows the limited range in ratios, we may understand the narrow range physically as being due to the very similar upper level temperatures of 55 K versus 62 K for CO(4-3) and \citwo, respectively. While their critical densities differ by a factor of about 20 and this is surely responsible for some of the difference seen in these ratios, the similar upper level temperatures are more important in these cases where density is not too far below critical.

However, the observed ratios exhibit an even narrower range than that populated by these `reasonable' models. The full range of observed ratios is shown by the grey band in all of these plots. So even if the models alone suggest a narrow range for the CO(4-3)/\citwo ratio, we find that real regions within galaxies have an even narrower range in this ratio, suggesting that all of our model parameter space is not explored by real galaxies.

For \cione with its 24 K upper level temperature and lower critical density, the narrowest range of ratio should occur for CO(1-0)/\cione based purely on the reasonable range models. But at least for the regions with \cione detections, the analysis summarized by Fig.~5 showed minimal scatter for the relationships between \cione and  CO(2-1), CO(3-2) or CO(4-3). This may be due to our detections in \cione being biased towards the brightest (likely higher-excitation) regions of these spiral galaxies or the lower quality of the ground-based CO(1-0) data.

\section{The \ci to \htwo conversion factor}

\begin{figure*}
\begin{center}
\includegraphics[width=16cm]{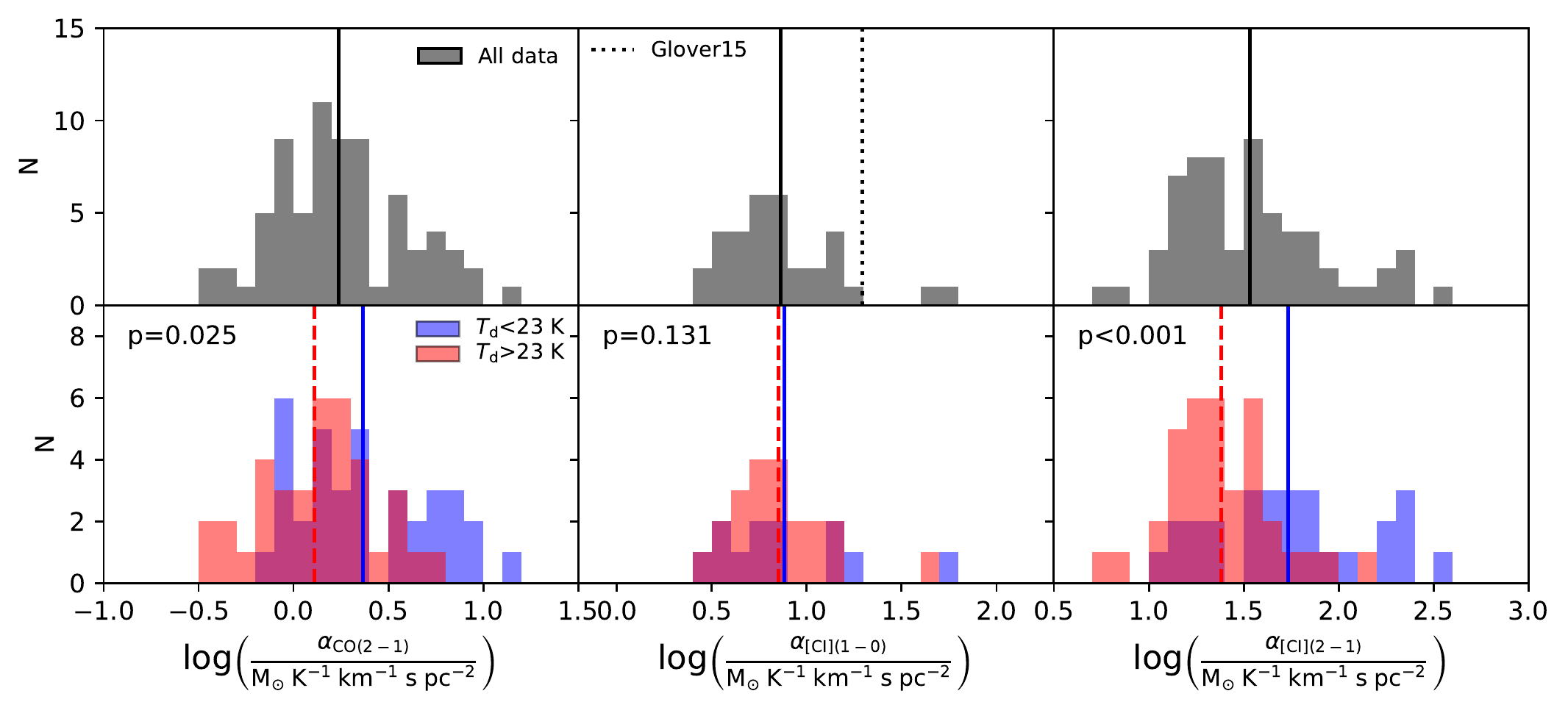}
\end{center}
\caption{Distributions of \alphaCOtwo, \alphaCIone and \alphaCItwo values for regions within BtP sample galaxies. The top row shows the full distributions while the bottom row's colored histograms break the distributions into two bins, based on a dust temperature of 23 K.  Probabilities that the two distributions are drawn from the same sample according to a two-sample Kolmogorov-Smirnov test are given in the upper left-hand corner of each subplot. The dashed red line gives the median of the warm-dust $\alpha$ values while the blue solid line gives the median of the cold-dust $\alpha$ values. The dotted line in the top center subplot indicates the \alphaCItwo value estimated from the \citet{glover15} simulation. \alphaCOtwo values are from \citet{sandstrom13}.}
\label{fig:alpha_CI}
\end{figure*}

One of the primary interests in the \ci lines is for their ability to measure the cold molecular gas mass at high redshift, where they are observationally easier to measure than the low-$J$ transitions of CO. Matching the definition of \alphaCO, we define 
\begin{equation}
\alpha_{\mathrm{[CI]}} = \frac{\Sigma_{\mathrm{mol}}}{I_{\mathrm{[CI]}}}
\end{equation}
where $\Sigma_{\mathrm{mol}}$ is the mass surface density of molecular gas (including helium) in M$_{\odot}$ pc$^{-2}$ and $I_{\mathrm{[CI]}}$ is the \ci line intensity in K km s$^{-1}$.  For most of the galaxies in the BtP sample, \alphaCOtwo values have been obtained in \citet{sandstrom13} using a technique that minimizes the spread in dust-to-gas ratio within a limited region using the observed dust, \hi and CO emission. We apply these values to calculate $\Sigma_{\mathrm {mol}}$ for the regions with detected \ci lines. Then, \alphaCI  may be estimated by dividing by the region's \ci intensity:
\begin{equation}
\alpha_{\mathrm{[CI]}} =\frac{\alpha_{\mathrm{CO(2-1)}} I_{\mathrm{CO(2-1)}}}{I_{\mathrm{[CI]}}}.
\end{equation}

The test here is to see how the distribution of \alphaCI compares to that of \alphaCO. Unfortunately, observational uncertainties mean we cannot compare the true distributions of these values, only their observed distributions. However, similar to the method used to calculate the intrinsic scatter about the power-law fits in Fig.~\ref{fig:ci_vs_co}, we can estimate the intrinsic spread of the distribution for \alphaCOtwo, \alphaCIone and \alphaCItwo since we know the uncertainties on the quantities used to compute them. We assume a lognormal distribution for these quantities.  We allow for both the observational uncertainties and an intrinsic spread of the distribution ($\sigma_{\mathrm{int}}$). The estimated $\sigma_{\mathrm{int}}$ (in dex) is that which gives a reduced $\chi^{2}$ value of 1:
\begin{equation}
\frac{\chi^2}{\nu} = \sum \frac{(\log_{10}{\mu} - \log_{10}{\alpha_{i}})^2}{\sigma_{i}^2 + \sigma_{\mathrm{int}}^2} = 1,
\end{equation}
where $\nu$ is the number of degrees of freedom, $\mu$ is the geometric mean of all the $\alpha_{i}$, $\alpha_{i}$ are the calculated values of $\alpha$ for each region, and $\sigma_{i}$ is the total observational uncertainty for each region, expressed in dex.
The \alphaCOtwo values have a typical uncertainty of about 0.2 dex \citep{sandstrom13}. The \alphaCI values have additional uncertainty from the measurements of  $I_{\mathrm{CO(2-1)}}$ and $I_{\mathrm{[CI]}}$ that go into computing them.

The top row of Fig.~\ref{fig:alpha_CI} shows the distributions of \alphaCOtwo, \alphaCIone and \alphaCItwo values for regions within BtP sample galaxies. Table~\ref{tab:alpha} tabulates the geometric mean $\alpha_{\mathrm{[CI]}}$ and \alphaCOtwo values and both the observed and intrinsic spreads expressed in dex along with their errors. There are two rows for \alphaCOtwo, because it must be constrained to the same set of data as its \alphaCI pair (fewer regions are available for \cione) for a fair comparison. Comparing these data-matched pairs, the spreads of \alphaCIone and \alphaCOtwo are not significantly different, while the spread of \alphaCItwo is greater than that of \alphaCOtwo. 
Thus \cione is potentially as good a tracer of the cold molecular gas as CO(2-1). On the other hand, \citwo appears to be a slightly worse tracer, in agreement with our finding of its good correlation with CO(4-3).

The factor \alphaCO is known to vary based on local excitation conditions. In particular, it has long been known that more highly excited galaxies like LIRGs and ULIRGs tend to have lower \alphaCO ratios (less molecular mass per unit emission). In the bottom panel of Fig.~\ref{fig:alpha_CI} we use the dust temperature to separate more and less excited regions within our sample galaxies. We divided the available data into two bins using a division of 23 K (the average dust temperature). Blue represents lower dust temperatures and red higher. The $p$-values listed in the upper left hand corner correspond to the probability that both distributions are samples drawn from the same parent distribution according to a Kolmogorov-Smirnov test. Values under 0.05 can be considered significant, so \alphaCOtwo and \alphaCItwo both show significance. In both of these cases, we see the expected pattern that higher-excitation regions tend to lower $\alpha$ values. The \alphaCIone distribution does not show a significant difference between low and high dust temperature regions, signaling perhaps it is a better molecular gas tracer because it is less dependent on excitation conditions. 

Our \alphaCIone values are mostly lower (our median value is about a factor of 3
    lower)  than the simulation-based predictions of $\alpha_{\mathrm{[CI](1-0)}} \approx 20$ M$_{\sun}$ K$^{-1}$ km$^{-1}$ s pc$^{-2}$ \citep[][dotted line in Fig.~\ref{fig:alpha_CI}]{offner14, glover15}. We suspect this difference occurs because these simulations do not trace the clouds after star formation begins and thus contain purely quiescent gas instead of the mix contained in our large apertures within disk galaxies. We also note that the regions with detected \ci emission have lower than typical \alphaCO values found on average for disks. When converted to the more common \alphaCOone via the \citet{sandstrom13} assumed $R_{21} = 0.7$, the mean value of \alphaCOone is 0.9 (or 1.1)  K$^{-1}$ km$^{-1}$ s pc$^{-2}$ for our regions detected in \cione (or \citwo). These values are both lower than the mean \alphaCOone = 3.1 M$_{\sun}$ K$^{-1}$ km$^{-1}$ s pc$^{-2}$  found by \citet{sandstrom13} whose sample contains many of these same galaxies, but probes further out in their disks.
However, as expected, the mean $\alpha_{\mathrm{[CI]}}$ values we calculate are larger than the \alphaCIone = 4.9 M$_{\sun}$ K$^{-1}$ km$^{-1}$ s pc$^{-2}$ and \alphaCItwo = 17 M$_{\sun}$ K$^{-1}$ km$^{-1}$ s pc$^{-2}$ found for (U)LIRGs by \citet{jiao17} by using a conversion of CO to $\mathrm{H}_2$ mass. On the other hand, \citet{jiao17} find similar values of \alphaCIone = 10.3 M$_{\sun}$ K$^{-1}$ km$^{-1}$ s pc$^{-2}$ and \alphaCItwo = 37.4 M$_{\sun}$ K$^{-1}$ km$^{-1}$ s pc$^{-2}$ when using a different method of inferring the $\mathrm{H}_2$ mass directly from the two \ci \xspace  lines, assuming they are optically thin and in LTE. (Note these are different than the values in \citet{jiao17}, adjusted by a factor 1.36 to include the contribution of helium to the total molecular gas mass.)

\begin{table}[htp]

\begin{center}
\caption{H$_{2}$ conversion factors for \ci detected regions.}
\label{tab:alpha}
\begin{tabular}{lccc}
\tableline

 & Mean & $\sigma_\mathrm{obs}$ (dex) & $\sigma_{\mathrm{int}}$ (dex)  \\
\tableline
$\alpha_{\mathrm{[CI](1-0)}} $& 7.3 & 0.29 & 0.20$^{+0.06}_{-0.05}$  \\
$\alpha_{\mathrm{CO(2-1)}} $& 1.30 & 0.28 & 0.21 $^{+0.06}_{-0.04}$ \\
\tableline
$\alpha_{\mathrm{[CI](2-1)}} $ & 34 & 0.39 & 0.32$^{+0.05}_{-0.04}$ \\
$\alpha_{\mathrm{CO(2-1)}} $& 1.56 & 0.33 & 0.27$^{+0.04}_{-0.03}$  \\

\tableline

\end{tabular}
\end{center}

\tablecomments{Units for the means are M$_{\sun}$ K$^{-1}$ km$^{-1}$ s pc$^{-2}$.  The two rows for \alphaCOtwo describe the \alphaCOtwo distribution constrained to the same set of regions as have detections in \cione and \citwo, respectively.}
\end{table}

\section{Conclusion}

We used spatially resolved Herschel SPIRE/FTS FIR spectroscopy from the Beyond the Peak program to measure \ci and CO line intensities in a sample of 18 nearby, normal galaxies. We found a linear relationship between the \citwo and CO(4-3) lines. This relationship also had minimal scatter compared to the relationships found with other CO lines. A simple physical explanation is the small difference between the upper level temperatures of the \citwo and CO(4-3) lines, which appears to be more important than their different critical densities.  Other relationships between \ci and CO lines are all sublinear (see Figs. 4 and 5). 

With a two-component LVG + high excitation template model, we find that \cione and CO(1-0) are dominated by the low-excitation gas, while CO(6-5) and CO(7-6) are dominated by the high-excitation component. \citwo, CO(2-1), CO(3-2), CO(4-3) and CO(5-4) have significant contributions from both low and high excitation components, but critically depend on local conditions.  

 We also determined \ci to molecular mass conversion factors for both \ci lines, with mean values of $\alpha_{\mathrm{[CI](1-0)}} = 7.3 $ M$_{\sun}$ K$^{-1}$ km$^{-1}$ s pc$^{-2}$ and $\alpha_{\mathrm{[CI](2-1)}} = 34 $ M$_{\sun}$ K$^{-1}$ km$^{-1}$ s pc$^{-2}$. \alphaCIone has a similar intrinsic spread (0.20$^{+0.06}_{-0.05}$ dex) to an \alphaCO value based upon CO(2-1) (0.21 $^{+0.06}_{-0.04}$ dex), thus we conclude that it is a similarly good tracer of the cold molecular gas.  However, \alphaCItwo has a larger intrinsic spread (0.32$^{+0.05}_{-0.04}$ dex) than \alphaCOtwo (0.27$^{+0.04}_{-0.03}$), signaling that much of the \citwo emission may originate in warmer molecular gas.

\acknowledgments{
CDW acknowledges support from the Natural Sciences and Engineering Research Council of Canada. MGW was supported in part by NSF grant AST1411827. EB acknowledges support from the UK Science and Technology Facilities Council [Grant No. ST/M001008/1]. LKH is grateful to funding by the INAF PRIN-SKA program 1.05.01.88.04. DR acknowledges support from the UK Science and Technology Facilities Council (Grant No. ST/N000919/1). ER acknowledges the support of the Natural Sciences and Engineering Research Council of Canada (NSERC), funding reference number RGPIN-2017-03987.

The Herschel spacecraft was designed, built, tested, and launched under a contract to ESA managed by the Herschel/Planck Project team by an industrial consortium under the overall responsibility of the prime contractor Thales Alenia Space (Cannes), and including Astrium (Friedrichshafen) responsible for the payload module and for system testing at spacecraft level, Thales Alenia Space (Turin) responsible for the service module, and Astrium (Toulouse) responsible for the telescope, with in excess of a hundred subcontractors. SPIRE has been developed by a consortium of institutes led by Cardiff University (UK) and including Univ. Lethbridge (Canada); NAOC (China); CEA, LAM (France); IFSI, Univ. Padua (Italy); IAC (Spain); Stockholm Observatory (Sweden); Imperial College London, RAL, UCL-MSSL, UKATC, Univ. Sussex (UK); and Caltech, JPL, NHSC, Univ. Colorado (USA). This development has been supported by national funding agencies: CSA (Canada); NAOC (China); CEA, CNES, CNRS (France); ASI (Italy); MCINN (Spain); SNSB (Sweden); STFC, UKSA (UK); and NASA (USA).
}

\software{RADEX (van der Tak et al. 2007), NADA (v1.6-1; Lee 2017), SciPy (http://www.scipy.org/)}

\bibliographystyle{apj}
\bibliography{master-apj}

\end{document}